\documentclass[prb,aps,twocolumn,showpacs,superscriptaddress]{revtex4}

\usepackage{graphicx} 
\usepackage{hyperref}
\usepackage{amsfonts}
\usepackage{amsmath,amssymb}
\usepackage{bm} 
\usepackage{color}
\usepackage{epstopdf} 
\usepackage{epsfig}
\usepackage{subfig} 
 \usepackage{float}

\def\be{\begin{equation}} 
\def\ee{\end{equation}}  
\def\bea{\begin{eqnarray}}
\def\eea{\end{eqnarray}}
\def\ba{\begin{array}}
\def\ea{\end{array}}
\def\bem{\begin{multline}} 
\def\eem{\end{multline}}

\begin{document}

\title{
Role of inter-edge tunneling in localizing Majorana zero modes at the ends of quasi one-dimensional $p+ip$ systems}
\author{Jimmy A. Hutasoit}
\email{jah77@psu.edu} 
\affiliation{Department of Physics, The Pennsylvania State University, University Park, Pennsylvania 16802, USA}
\author{Ajit C. Balram}
\email{ajit@phys.psu.edu} 
\affiliation{Department of Physics, The Pennsylvania State University, University Park, Pennsylvania 16802, USA}
\date{\today}

\begin{abstract} Potter and Lee have demonstrated the presence of Majorana zero modes at the ends of quasi one-dimensional (1-D) $p+ip$ superconductors \cite{Potter:2010ve}. Here, we use conformal field theory (CFT) methods to show that inter-edge tunneling of the vortex excitations along the length of the channel is crucial for such localization. We then show that localization of Majorana modes occurs also in quasi 1-D channels of the $5/2$  fractional quantum Hall (FQH) systems when modeled, following Moore and Read, as the $p+ip$ paired state of composite fermions. Finally, we propose a tunnel-interferometry experiment to detect these modes, which should show a $\pi$ phase shift of oscillations depending on whether or not a localized Majorana zero mode is present, which, in turn, can be controlled by varying the tunneling strength. 
\end{abstract}

\pacs{71.10.Pm, 74.78.-w, 73.43.Jn, 03.67.Lx}

\maketitle

\section{Introduction}

Recently, the zero energy modes of Majorana fermions have been receiving a substantial amount of interest due to their potential applications in decoherence-free quantum computation \cite{Kitaev:qf} (for recent reviews, see for example, Refs.  \onlinecite{Beenakker:2011fk} and \onlinecite{Alicea:2012uq}). The proposals on how one can realize Majorana zero modes in condensed matter systems include a superconductor--strong topological insulator heterostructure \cite{Fu:2008kx} and semiconductor--superconductor heterostructures \cite{Sau:2010vn,Lutchyn:2010ys}. The basic idea of these proposals is to produce 1- or 2-D $p+ip$ superconductors, wherein the Majorana zero modes will be localized either at the ends of the wire \cite{Kitaev:qf} or in the vortex cores \cite{Read:2000zr,Ivanov:ly}.  Furthermore, numerical calculations show that Majorana zero modes at the ends of the $p+ip$ superconducting wire are robust beyond the strict 1-D limit as long as the sample width is not much larger than the superconducting coherence length \cite{Potter:2010ve}. 

In this article, we show, using CFT methods, that the crucial ingredient in such localization of Majorana zero modes at the ends of quasi 1-D $p+ip$ superconductors is the inter-edge tunneling, which is dominated, in a renormalization group (RG) sense, by the tunneling of the vortex modes
. This CFT treatment not only provides an additional physical insight into the numerical results of Ref. \onlinecite{Potter:2010ve} but it also provides us with a concrete way to argue that Majorana zero modes also exist at the ends of quasi 1-D $\nu =5/2$ FQH systems. This is due to the fact that the edge theory of a $p+ip$ superconductor also describes, according to the Moore-Read model, the neutral sector of the $\nu =5/2$ FQH system. 

This is a welcomed development for several reasons. From the theoretical point of view, unlike the case of $p+ip$ superconductors, the quasi 1-D $\nu=5/2$ FQHE systems have thus far not been numerically tractable. Therefore, even though conceptually, the $\nu=5/2$ FQHE system is connected to the $p+ip$ superconductor via composite fermion theory, prior to this work, there has not been a method to determine whether the quasi 1-D $\nu=5/2$ FQHE system also exhibits Majorana zero modes at its ends. From the experimental point of view, the wider availability of materials that actualize $\nu=5/2$ FQH states means that there are more platforms to realize Majorana zero modes. Furthermore, this might lead to a wider experimental window for detecting and manipulating these zero modes. Here, we propose one such experiment where a geometry with variable width may be used to detect such modes through a $\pi$ phase shift in an interference experiment. 

\section{Quasi 1-D $p+ip$ Superconductor}

\begin{figure}
\begin{center} 
\includegraphics[width=2.5 in]{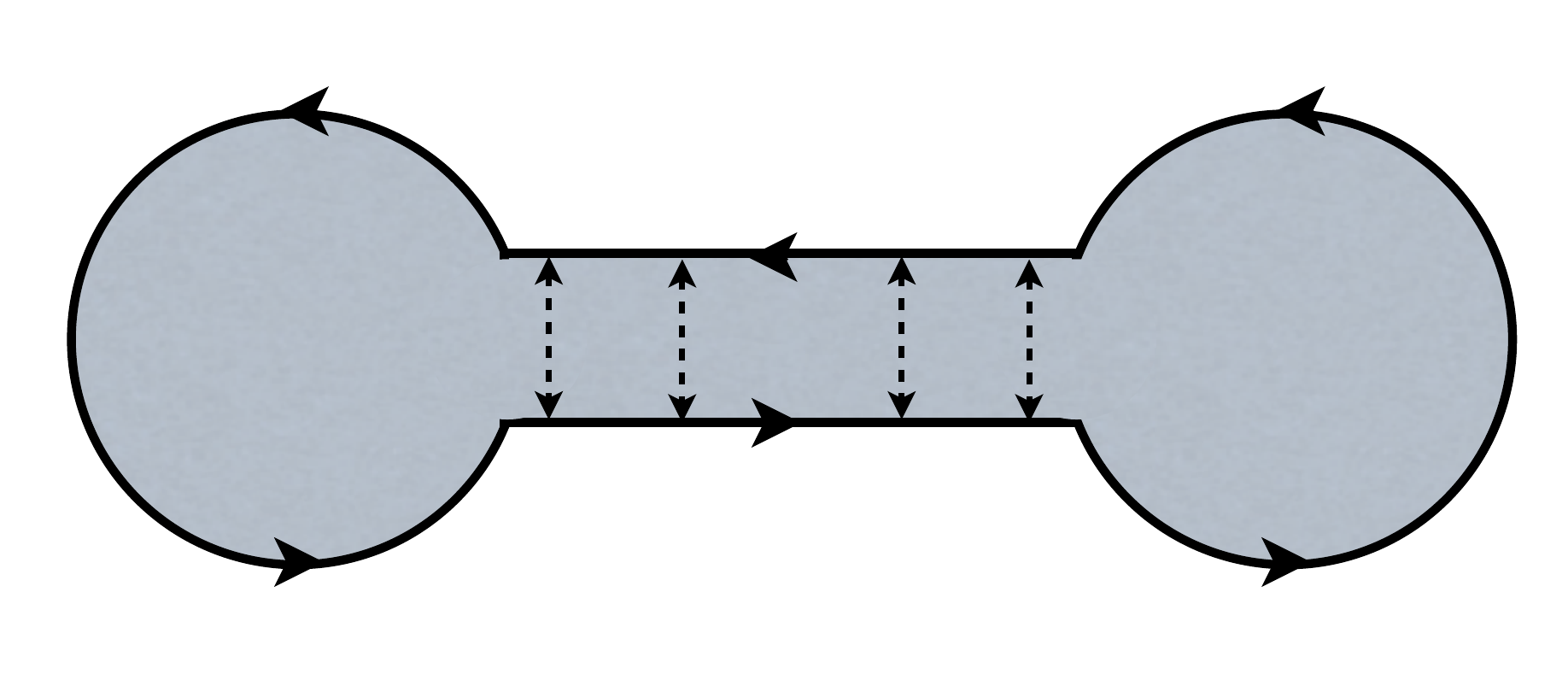}
\end{center}  
\caption{The dumbbell-like geometry of the $p+ip$ superconductor. The solid line represents the chiral edge states and the dashed lines represent the tunneling of the edge excitations across the constriction.  \label{fig:dumbell} }
\end{figure}

Let us start by first considering the $p+ip$ superconductor. We find it convenient to consider a dumbbell-like geometry as depicted in Fig. \ref{fig:dumbell}. The key idea is to have a long constriction whose width is not much larger than the superconducting coherence length, such that the edge excitations can tunnel across this constriction\footnote{This is a necessary but not sufficient condition for such tunneling to occur, as the presence of disorder is also needed.}. Such a constriction can be obtained by directly manufacturing a long strip with small width or by applying a gate voltage on either sides of the superconductor \cite{Fendley:2007fk}. Allowing the edge excitations to tunnel across the superfluid will be seen as crucial in producing Majorana zero modes at the end of the constriction. Inter-edge tunneling will not only open a gap at the constriction, but also cause the lowest energy configuration to contain vortices at the ends, implying localized Majorana zero modes at the ends of the constriction.

The edge theory of a $p+ip$ superconductor is described by a deformed chiral Ising CFT and, focusing  on the constriction part, the (Euclidean) action is given by
\begin{eqnarray}
S &=& \int dx\, d\tau \left( \mathcal{L}_{\rm L} + \mathcal{L}_{\rm R} +\mathcal{L}_{\rm tun} \right),\\
\mathcal{L}_{\rm L} &=& \psi_L \left(\partial_{\tau} + iv\, \partial_x\right) \psi_L,\\
\mathcal{L}_{\rm R} &=&  \psi_R \left(\partial_{\tau} - iv\, \partial_x\right) \psi_R,\\
\mathcal{L}_{\rm tun} &=& i\, \lambda_{\psi} \,\psi_L \psi_R + \lambda_{\sigma} \,\sigma_L \sigma_R, \label{eq:tunsc}
\end{eqnarray}
where $\mathcal{L}_{\rm L}$ and $\mathcal{L}_{\rm R}$ refer to left and right moving Majorana modes, respectively, the ``deforming" term $\mathcal{L}_{\rm tun}$ describes the tunneling, 
and $\tau = i t$ is the imaginary time. The conformal dimensions of the fields are 
\be
[\psi_{L,R}]=\frac{1}{2} \quad {\rm and} \quad  [\sigma_{L,R}]=\frac{1}{16}.
\ee
Here, $\sigma_{L,R}$ are chiral spin fields\footnote{Loosely speaking, the chiral spin field,  which corresponds to a vortex, can be thought of as the ``square root" of the non-chiral spin field. This statement is not exact due to the fact that the non-chiral spin field does not factorize into the product of chiral spin fields \cite{Fendley:2007fk}.}. 
 
The tunneling operators in Eq. \ref{eq:tunsc} are relevant in the RG sense and the RG flow is dictated by 
\bea
\frac{d}{dl}  \lambda_{\psi} =  \lambda_{\psi}, \qquad \frac{d}{dl}  \lambda_{\sigma} = \frac{15}{8} \lambda_{\sigma}.
\eea
These RG equations are not valid up to arbitrary strong coupling as one expects the ``weak" tunneling of Eq. \ref{eq:tunsc} will become unstable and will result in breaking the superfluid into multiple weakly connected droplets \cite{Fendley:2007fk}.


In this regime of ``weak" tunneling, the tunneling term involving spin fields is more relevant than the one involving Majorana fields and thus, $\lambda_{\psi} \ll \lambda_{\sigma}$. Furthermore, from the first RG equation, we see that the less relevant term, which can also be thought of as a mass term, results in opening a gap along the constriction because $\lambda_{\psi}$ grows as one flows to the infrared. 

\begin{figure*}
\begin{center} 
\includegraphics[width= 5 in ]{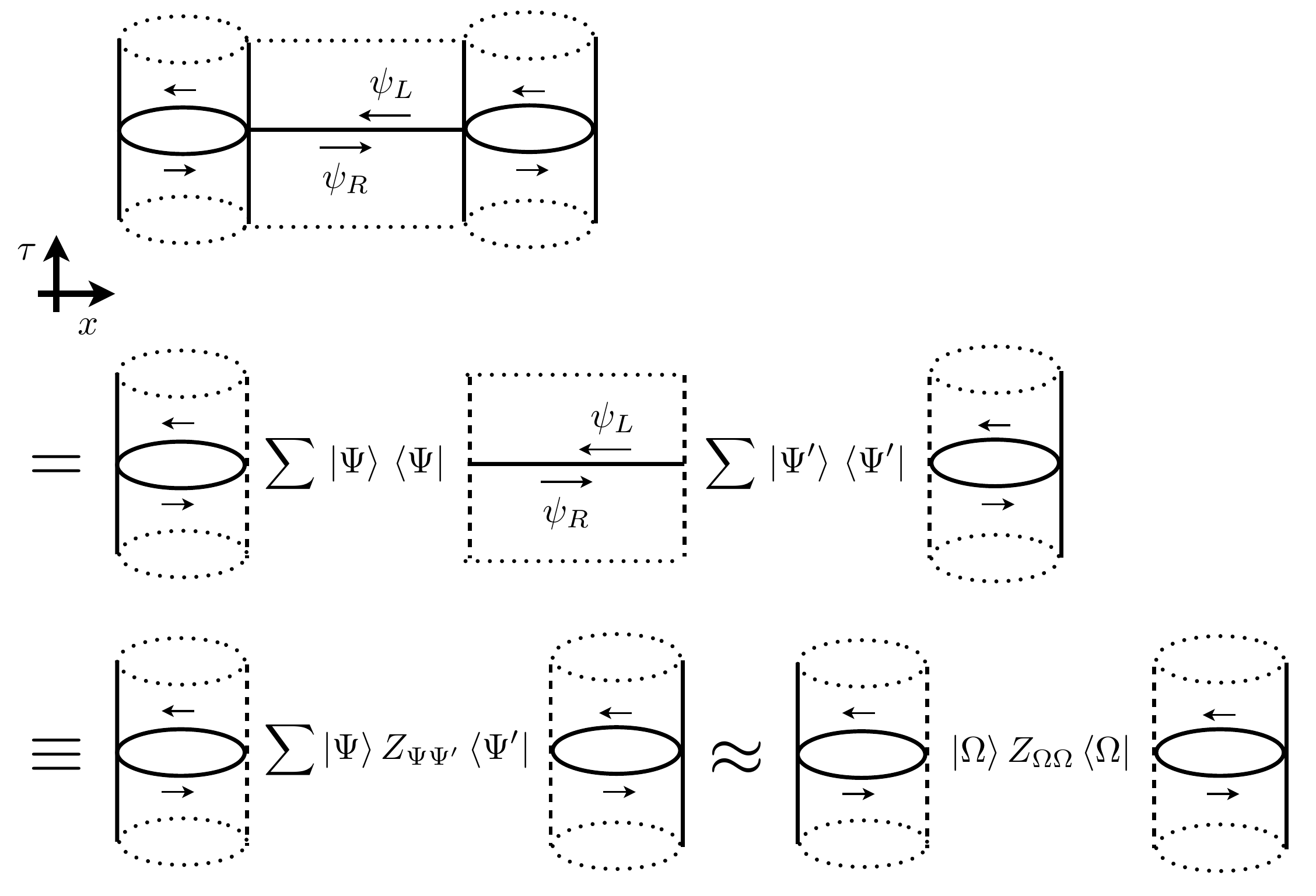}
\end{center}  
\caption{The edge theory lives in a (1+1)-D spacetime that looks like a strip sandwiched by two cylinders. Here, the vertical direction is the temporal direction. Using the sewing property of (1+1)-D CFT, we factorize the geometry by inserting complete sets of states at the cuts, which are represented by the dashed lines. The middle part is the transfer matrix element along the strip $Z_{\Psi \Psi'}$ and for long constriction, the sum over all states is dominated by the state $|\Omega\rangle$ whose energy is the lowest in the deformed CFT.   \label{fig:sew} }
\end{figure*}

To understand the effect of the more relevant term, it is instructive to continuously deform the (1+1)-D manifold of the edge theory into the geometry depicted on the first line of Fig. \ref{fig:sew}. Using the sewing property of (1+1)-D CFT, we obtain the second line of Fig. \ref{fig:sew}, where we have factorized the manifold by inserting complete sets of states at the cuts\footnote{Such property has been used to calculate higher order amplitudes in perturbative string theory, see for example Ch. 9 of Ref. \onlinecite{Polchinski:1998rq}.}. The middle part, which corresponds to the constriction, is nothing but the transfer matrix element along the strip, which in diagonal basis is given by 
\be
Z_{\Psi_d \Psi'_d} = \delta_{\Psi_d \Psi_d'} \, \exp\left(-L H_{\Psi_d}\right), 
\ee
where $L$ is the length of the constriction and the spectrum $H_{\Psi_d}$ is obtained by diagonalizing the Hamiltonian \cite{Zamolodchikov:1989zs,Yurov:1989yu,Lassig:1990xy,Lassig:1990cq}. If $L\gg1$ is the largest length scale in the system, then the sum over the states is dominated by the highest weight state $|\Omega \rangle$ of the deformed CFT.

When there is no tunneling, the edge theory at the constriction is just the usual Ising CFT and thus, 
\be
|\Omega\rangle = |0\rangle,
\ee
where $|0\rangle$ is its vacuum state. However, as we will see below, tunneling causes $|\Omega\rangle$ to be a linear combination of the Ising CFT vacuum state $|0\rangle$ and the so-called spin state $|\sigma\rangle$, which contains a vortex. 

To obtain $|\Omega \rangle$, we need to calculate the spectrum of the deformed CFT and find the state with the lowest energy. The spectrum is obtained by diagonalizing\footnote{We assume periodic boundary condition along the temporal direction $\tau \to \tau + 2 \pi$.} 
\bea
\langle \Psi | H | \Psi' \rangle &=& \langle \Psi | \left(L_0 + \bar{\rm L}_0 - \frac{c}{12}\right) | \Psi' \rangle  +  \lambda_{\sigma} \langle \Psi | \sigma \bar{\sigma} | \Psi' \rangle, \nonumber\\
\eea
where $\Psi$ and $\Psi'$ are the states of the undeformed CFT, $L_0$ and $\bar{\rm L}_0$ are the Virasoro generators, and $c$ is the central charge, which for the Ising CFT equals 1/2. Here, following Ref.  \onlinecite{Zamolodchikov:1989zs}, we have kept only the most relevant term that deforms the theory away from the conformal fixed point. 

Due to the integrability of a certain class of unitary CFTs, which includes Ising CFT, the above statement is not a perturbative statement \cite{Zamolodchikov:1989zs} and is valid even for large $\lambda_{\sigma}$. Instead, the accuracy of the spectrum obtained depends on how many states one includes before diagonalizing the matrix $\langle \Psi | H | \Psi' \rangle$ \cite{Yurov:1989yu,Lassig:1990xy,Lassig:1990cq}. However, since we are interested in knowing only what the state with the lowest energy is, it suffices to include only the primary states of the Ising CFT: $|0\rangle$, $|\sigma\rangle$ and $|\epsilon\rangle$. Similarly, the small corrections due to the less relevant tunneling term will not affect our qualitative results. Here, $|\sigma\rangle$ and $|\epsilon\rangle$ are defined as per the usual radial quantization procedure \cite{francesco1996conformal} as
\be
|\sigma\rangle = \sigma_{nc}(z=0, \bar{z}=0) |0\rangle, \qquad |\epsilon\rangle = \epsilon(z=0, \bar{z}=0) |0\rangle,
\ee
where $\sigma_{nc}(z, \bar{z})$ and $\epsilon(z, \bar{z})$ are the non-chiral spin field and energy field, respectively. We have also introduced the complex coordinate 
\be
z = \exp \left(- x-i v \tau\right)
\ee
and its complex conjugate $\bar{z}$. The energy field is the product of the chiral Majorana fields 
\be
\epsilon(z, \bar{z}) = i \bar{\psi} (\bar{z}) \psi (z),
\ee 
while the non-chiral spin field does \emph{not} factorize into the product of the chiral spin fields \cite{Fendley:2007fk}. Here, we have also replaced the (left or right) chiral notation for the fields with holomorphic or anti-holomorphic notation. 

In the basis $\{|0\rangle, |\sigma\rangle, |\epsilon\rangle\}$, we have (see Appendix \ref{appx} for details)
\be
\langle \Psi | H | \Psi' \rangle =   \left( \begin{array}{ccc} 
      -\frac{1}{24} & \lambda_{\sigma} & 0\\
      \lambda_{\sigma} & \frac{1}{12} & \frac{\lambda_{\sigma}}{2} \\
     0&  \frac{\lambda_{\sigma}}{2}  & \frac{23}{24} 
   \end{array}\right)   ,
\ee   
and
\bea
|\Omega\rangle &=&  |0\rangle - 8\, \lambda_{\sigma} \, |\sigma\rangle + {\cal O} \left(\lambda_{\sigma}^2\right). \label{eq:pipstate}
\eea

Since $|\Omega\rangle$ now contains $|\sigma\rangle$, the theories living on the cylinders (see Fig. \ref{fig:sew}) are chiral Ising CFTs, but with spin fields inserted at the cuts. The effect of the spin field on the Majorana field can be seen from the operator product expansion (OPE)
\be
\psi(w)\sigma_{nc}(z,\bar{z})=\frac{e^{i \pi/4}}{\sqrt{2}(w-z)^{1/2}}\,\mu(z,\bar{z}),
\ee
where the disorder field $\mu$ has the same dimension of $\sigma_{nc}$. From the above equation, we see that if we rotate $z$ around $w$ by an angle of $2\pi$ we pick a phase of 
\be
[\exp(\pm 2\pi i)]^{1/2}=-1.
\ee
Thus, the spin field introduces a square root branch cut in the fermion correlators. In other words, inserting the spin field $\sigma_{nc}(z,\bar{z})$ at a given point in spacetime changes the fermionic boundary conditions around the point from antiperiodic to periodic, and thereby introduces a Majorana zero mode that is localized at that point. Therefore, the long constriction exhibits localized Majorana zero modes at its ends, in agreement with Ref. \onlinecite{Potter:2010ve}. It is worth emphasizing that the crucial ingredient of this phenomenon is the inter-edge tunneling of the vortex modes.

Physically, the inter-edge tunneling of the vortex modes causes the lowest energy configuration to consist of vortices trapped at the ends of the constriction. This in turn implies that there are localized Majorana zero modes at the ends of the constriction. The typical localization length of these zero modes is then equivalent to the typical vortex size which is of the order of the superconducting coherence length. 

\section{$\nu = 5/2$ Fractional Quantum Hall System}

Armed with the understanding of the case of the $p+ip$ superconductor, we can now apply the above CFT methods to the case of Moore-Read state in the $\nu = 5/2$ FQH system. In this case, the edge theory is described by a CFT which consists of chiral Ising CFT and a scalar field $\varphi$. Again, focusing only on the constriction part, the (Euclidean) action is given by
\begin{equation}
S = \int dx\, d\tau \left( \mathcal{L}_{\rm L} + \mathcal{L}_{\rm R} +\mathcal{L}_{\rm tun} \right),
\end{equation}
\be
\mathcal{L}_{\rm L} =  \psi_L \left(\partial_{\tau} + iv_n\, \partial_x\right) \psi_L + \frac{1}{2\pi} \partial_x \varphi_L \left(\partial_{\tau} + v_c \partial_x\right) \varphi_L,
\ee
\be
\mathcal{L}_{\rm R} =  \psi_R \left(\partial_{\tau} - iv_n\, \partial_x\right) \psi_R + \frac{1}{2\pi} \partial_x \varphi_R \left(\partial_{\tau}- v_c \partial_x\right) \varphi_R,
\ee
\be
\mathcal{L}_{\rm tun} = i\, \lambda_{\psi} \,\psi_L \psi_R + \lambda_{\varphi} \cos \frac{\varphi_L - \varphi_R}{\sqrt{2}}+ \lambda_{\sigma} \,\sigma_L \sigma_R \cos \frac{\varphi_L - \varphi_R}{2\sqrt{2}}.
\ee
Here, 
\be
[\exp(i\varphi_{L,R})]=\frac{1}{2},
\ee
and thus, the RG flow of this theory is given by 
\bea
\frac{d}{dl}  \lambda_{\psi} =  \lambda_{\psi}, \qquad\frac{d}{dl} \lambda_{\varphi} = \frac{3}{2} \lambda_{\varphi}, \qquad  \frac{d}{dl}  \lambda_{\sigma} = \frac{7}{4} \lambda_{\sigma}.
\eea
As is in the case of the $p+ip$ superconductor, the tunneling term involving chiral spin fields is the most relevant term\footnote{This term corresponds to the tunneling of quasiholes.}. Moreover, from the first two RG equations, we see that the other tunneling terms result in opening gaps for the edge modes, both the charged scalar modes and the Majorana modes. Just like in the case of the $p+ip$ superconductor, the most relevant tunneling term in this case will also result in the insertion of non-chiral spin operators at the end parts of the geometry. 

The primary states of the scalar sector of the theory that are relevant to our problem are given by
\be
|\pm \varphi \rangle = :\exp \left(\pm \, i \, \frac{\varphi_L - \varphi_R}{2\sqrt{2}} \right):\Bigg|_{z=0} |0\rangle,
\ee
where $:\,\, :$ denote the normal ordering. In the basis $\{ |-\varphi\rangle, |0\rangle, |\sigma \rangle, |\epsilon \rangle,    |\varphi\rangle\}$, we have
\be
\langle \Psi | H | \Psi' \rangle =   \left( \begin{array}{ccccc} 
      \frac{1}{32\pi^2}-\frac{1}{8}&0 &\frac{ \lambda_{\sigma}}{2} & 0& 0\\
       0&-\frac{1}{8} & \lambda_{\sigma} & 0&0\\
      \frac{ \lambda_{\sigma}}{2} & \lambda_{\sigma} & 0 &\frac{ \lambda_{\sigma}}{2} & \frac{ \lambda_{\sigma}}{2} \\
     0 &0&\frac{ \lambda_{\sigma}}{2} & \frac{7}{8}&0 \\
     0 &0&\frac{ \lambda_{\sigma}}{2} &0&  \frac{1}{32\pi^2}-\frac{1}{8}
   \end{array}\right)   ,
\ee   
and
\bea
|\Omega\rangle &=&  |0\rangle - 8\, \lambda_{\sigma} \, |\sigma\rangle + {\cal O} \left(\lambda_{\sigma}^2\right). \label{eq:5/2state}
\eea

Therefore, just as in the case of the $p+ip$ superconductor, 
the long constriction in a Moore-Read $\nu=5/2$ FQH system will also exhibit localized Majorana zero modes at its ends. In this case, the crucial ingredient for such localization is the inter-edge tunneling of the quasiholes.

\section{Summary and Discussions} 

Combining the results of the RG equations and the deformed CFT spectrum on a strip, we have shown that the tunneling of the edge excitations across the $p+ip$ superfluid or $\nu=5/2$ FQH liquid in the constriction not only opens a gap at the constriction, but also cause the lowest energy configuration to contain vortices at the ends of the geometry. This results in localized Majorana zero modes at the end parts of the geometry. 

\begin{figure}
\begin{center} 
\includegraphics[width= 2.5 in ]{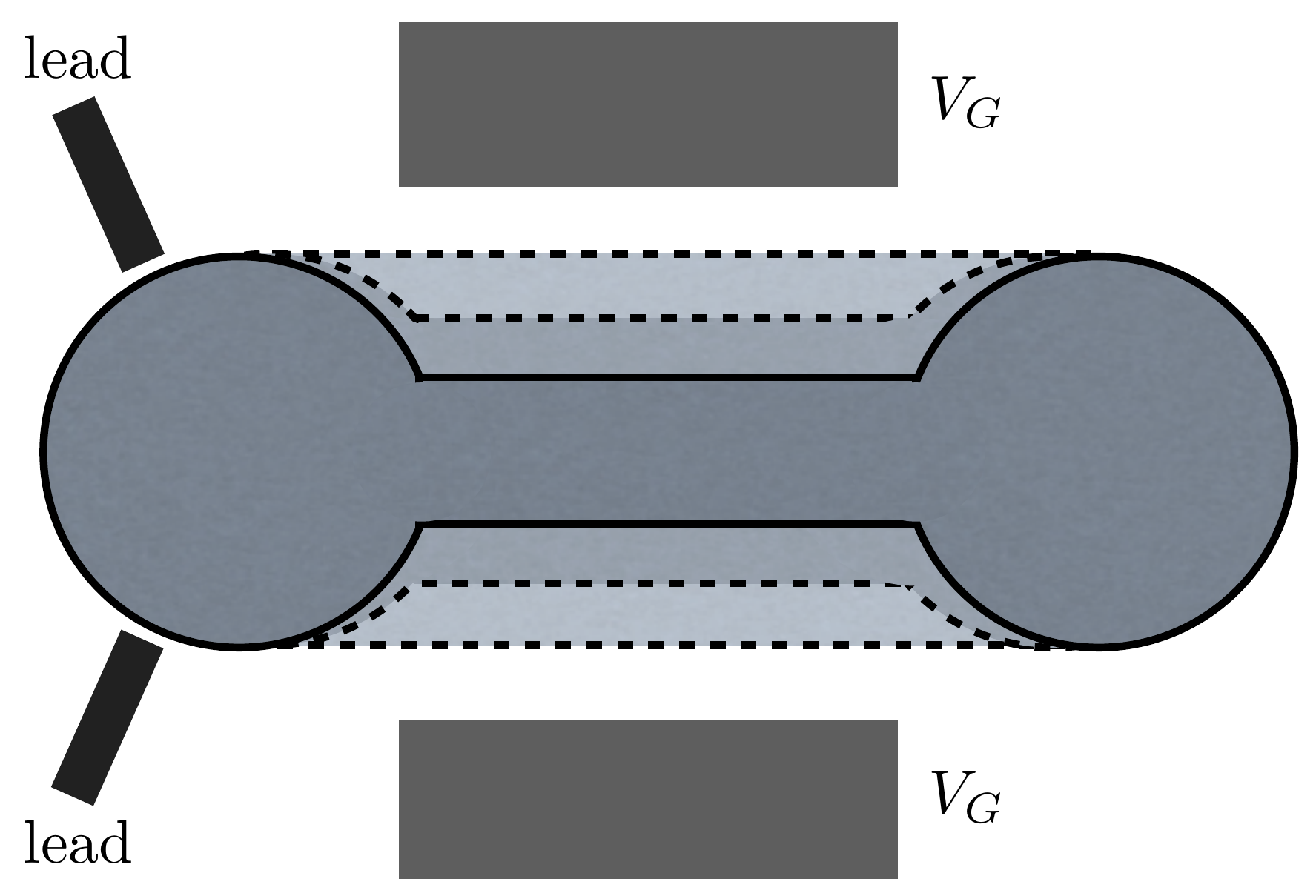}
\end{center}  
\caption{As one increases $V_G$, the FQH liquid is pinched to create the constriction (depicted as transitioning from the lighter shade to darker ones) and thus turns on Majorana zero modes at the ends. \label{fig:exp} }
\end{figure}

The emergence of these Majorana zero modes should be clearly visible once
\be
\lambda_{\sigma} \lesssim \frac{1}{8},
\ee
as can be seen from Eqs. \ref{eq:pipstate} and \ref{eq:5/2state}. In particular, for the case of the $\nu=5/2$ FQH system, we can observe these zero modes using an interferometry measurement as depicted in Fig. \ref{fig:exp}. Here, we assume a ``sharp" edge, where the electrons from the leads can tunnel all the way into the half-filled first excited Landau level. 

Let us consider a slab of $\nu=5/2$ FQH system with significant width and no applied gate voltage $V_G = 0$, so that there is no edge excitations tunneling. We then increase $V_G$ to create the constriction and thus, at a certain point, allow edge excitations to tunnel across the constriction. This results in having Majorana zero modes at the ends, which introduce a $\pi$ phase shift in the interferometric oscillation measured through the leads at the end part.

As we increase $V_G$ further, we will eventually cross into the ``strong" tunneling regime, where we pinch off the FQH liquid totally, splitting it into two decoupled FQH systems. At this point, there will be no chiral spin fields tunneling across  \cite{Fendley:2007fk} and therefore, there will be no Majorana zero modes at the ends. This results in a phase shift {\em back} to the original interferometric pattern. To summarize, starting at $V_G=0$, as we increase $V_G$, there will be a phase shift in the interferometric pattern at some point, which will then shift back as we keep on increasing $V_G$. 

A comment is in order. It is highly unlikely that one can increase the gate voltage while keeping the edges parallel. Instead, the edges will exhibit shallow parabolic profiles and the inter-edge tunneling strength will not be constant throughout the constriction. However, in the long constriction limit, the change in the tunneling strength will be mild. This is due to the fact that such change goes like the curvature 
\be
\frac{\partial_x \lambda_{\sigma}}{\lambda_{\sigma}} \sim \frac{1}{L},
\ee
and therefore, the correction to the spectrum due to this varying tunneling strength is of the order of ${\cal O}(1/L)$. Thus, our previous conclusion that the lowest energy configuration consists of two Majorana zero modes trapped at the ends of the constriction is still valid.

Lastly, let us note that we have treated the $\nu=5/2$ FQH state assuming the Moore-Read Pfaffian model, although the presence of Majorana edge modes from interacting microscopic theory is not yet conclusive \cite{Wan:2008fk}. An observation of localized Majorana modes will serve as a confirmation of certain non-trivial aspects of this model.

\acknowledgements

We thank Jainendra Jain, Radu Roiban, Paul Fendley, Xiao-Liang Qi, Chang-Yu Hou and Diptiman Sen for insightful discussions. This work is supported by NSF Grants No. DMR-1005536 and No. DMR-0820404 (Penn State MRSEC).

\appendix
\section{Matrix Elements of $\langle \Psi | H | \Psi' \rangle$} \label{appx}
The non-vanishing off-diagonal matrix elements $\langle \Psi | H | \Psi' \rangle$ for the case of the $p+ip$ superconductor are given by
\bea
\langle 0 |  \sigma \bar{\sigma} | \sigma \rangle &=& \langle \sigma(1) \bar{\sigma}(1) \sigma_{nc}(0,0) \rangle,\nonumber \\
\langle \sigma |  \sigma \bar{\sigma} | \epsilon \rangle &=& \lim_{z, \bar{z} \to 0} z^{-\tfrac{1}{8}} \bar{z}^{-\tfrac{1}{8}} \langle \sigma_{nc}(1/z,1/\bar{z}) \sigma(1) \bar{\sigma}(1) \epsilon(0,0) \rangle,\nonumber \\ \label{eq:offdiagonal}
\eea
and their complex conjugates. 
We can calculate these 2- and 3- point functions by introducing another copy of the theory, bosonizing and taking the ``square root" of the corresponding boson correlators \cite{DiFrancesco1987527}. The non-chiral spin field is bosonized such that  \cite{DiFrancesco1987527}
\bea
\langle \cdots \sigma_{nc}(z,\bar{z}) \cdots \rangle^2 &=& \langle \cdots \sigma^a_{nc}(z,\bar{z}) \cdots \rangle \langle \cdots \sigma^b_{nc}(z,\bar{z}) \cdots \rangle \nonumber \\
&=& N \left\langle \cdots \cos \frac{\phi(z,\bar{z})}{2} \cdots  \right\rangle,
\eea
where $a$, $b$ label the different copies of the theory, the scalar field
\be
\phi(z,\bar{z}) = \phi(z) - \bar{\phi}(\bar{z}), 
\ee
has the correlation function
\be
\langle \phi(z,\bar{z}) \phi(w,\bar{w}) \rangle = - \log |z-w|^2,
\ee 
and the normalization constant $N$ is chosen such that the correlation functions will exhibit the correct short distance behavior. For example, the 2-point function is given by
\be
\langle \sigma_{nc}(z,\bar{z}) \sigma_{nc}(w,\bar{w}) \rangle^2 = 2 \left\langle \cos \frac{\phi(z,\bar{z})}{2} \, \cos \frac{\phi(w,\bar{w})}{2}  \right\rangle.
\ee

Similarly, for the tunneling operator, we have 
\be 
\langle \cdots \sigma(z) \bar{\sigma} (\bar{z}) \cdots \rangle^2 = N \langle \cdots e^{\pm i \phi(z,\bar{z})/2} \cdots \rangle,
\ee
where the normalization constant $N$ is again chosen such that the correlation functions will exhibit the correct short distance behavior, while the relative signs between tunneling operators in the correlation functions are chosen to match the chiral conformal blocks \cite{Fendley:2007fk}. For example
\be
\langle \sigma(z) \bar{\sigma} (\bar{z}) \sigma(w) \bar{\sigma} (\bar{w}) \rangle^2 = \langle e^{i \phi(z,\bar{z})/2} e^{- i \phi(w,\bar{w})/2} \rangle,
\ee
in agreement with Refs. \onlinecite{Fendley:2007fk} and \onlinecite{Ardonne:2010fk}. We note that Refs. \onlinecite{Fendley:2007fk} and \onlinecite{Ardonne:2010fk} give the chiral correlation functions, from which one can deduce the anti-chiral correlation functions. In particular, 
\be
\langle \sigma(z) \bar{\sigma} (\bar{z}) \sigma(w) \bar{\sigma} (\bar{w}) \rangle^2
\ee 
is matched with 
\be
\langle \sigma_a(z) \sigma_b(z) \sigma_a(w) \sigma_b(w) \rangle \, \langle \bar{\sigma}_a( \bar{z})  \bar{\sigma}_b( \bar{z})  \bar{\sigma}_a( \bar{w})  \bar{\sigma}_b( \bar{w}) \rangle
\ee
of Ref.  \onlinecite{Fendley:2007fk}. 

We then have
\be
\langle \sigma(z) \bar{\sigma} (\bar{z}) \sigma_{nc} (w,\bar{w}) \rangle^2 = N \left \langle e^{i \phi(z,\bar{z})/2} \cos \frac{\phi(w,\bar{w})}{2}  \right\rangle.
\ee
We can evaluate $N$ from the operator product expansion (OPE) of 
\be
\sigma(z_1) \bar{\sigma}(\bar{z}_1) \sigma(z_2) \bar{\sigma}(\bar{z}_2) \sigma_{nc}(z_3,\bar{z}_3)  \sigma_{nc}(z_4,\bar{z}_4),
\ee 
which can be expressed as either 
\be
\Big[\sigma(z_1) \bar{\sigma}(\bar{z}_1) \sigma_{nc}(z_3,\bar{z}_3)\Big] \Big[\sigma(z_2) \bar{\sigma}(\bar{z}_2)   \sigma_{nc}(z_4,\bar{z}_4)\Big],
\ee 
or
\be
\Big[\sigma(z_1) \bar{\sigma}(\bar{z}_1) \sigma(z_2) \bar{\sigma}(\bar{z}_2)\Big] \Big[ \sigma_{nc}(z_3,\bar{z}_3)  \sigma_{nc}(z_4,\bar{z}_4)\Big].
\ee 
We then find $N=2$.

Therefore, we have
\bea
\langle 0 |  \sigma \bar{\sigma} | \sigma \rangle &=& \langle \sigma(1) \bar{\sigma}(1) \sigma_{nc}(0,0) \rangle = 1,\nonumber \\
\langle \sigma |  \sigma \bar{\sigma} | \epsilon \rangle &=& \lim_{z, \bar{z} \to 0} z^{-\tfrac{1}{8}} \bar{z}^{-\tfrac{1}{8}} \langle \sigma_{nc}(1/z,1/\bar{z}) \sigma(1) \bar{\sigma}(1) \epsilon(0,0) \rangle\nonumber \\ 
&=& \frac{1}{2}.
\eea

\bibliography{References} 

\end{document}